\documentclass[prd,preprint,aps,superscriptaddress,noshowpacs,preprintnumbers,nofootinbib]{revtex4}
\usepackage{varioref,exscale,latexsym,amsmath,amssymb,caption}
\usepackage[pdftex]{graphicx}
\usepackage{bm}
\usepackage{slashed}
\usepackage{epsfig}
\usepackage{color}
\usepackage[utf8]{inputenc}
\usepackage{feynmp}
\usepackage{subcaption}
\makeatletter
\newcommand*{\rom}[1]{\expandafter\@slowromancap\romannumeral #1@}
\makeatother
\begin{document}
\title{A new solar neutrino channel for grand-unification monopole searches}

\author{Nick Houston}\email{nhouston@itp.ac.cn}
\affiliation{CAS Key Laboratory of Theoretical Physics, Institute of Theoretical Physics, Chinese Academy of Sciences, Beijing 100190, P. R. China}
\author{Tianjun Li}\email{tli@itp.ac.cn}
\affiliation{CAS Key Laboratory of Theoretical Physics, Institute of Theoretical Physics, Chinese Academy of Sciences, Beijing 100190, P. R. China}
\author{Chen Sun}\email{chen.sun@dartmouth.edu}
\affiliation{CAS Key Laboratory of Theoretical Physics, Institute of Theoretical Physics, Chinese Academy of Sciences, Beijing 100190, P. R. China}
\affiliation{Department of Physics and Astronomy, Dartmouth College,
  Hanover, NH 03755, USA}

\date{\today}

\begin{abstract}
  We identify a previously
  untapped discovery channel for grand-unification monopoles,
  arising from their ability to catalyse the direct decay of protons
  into monoenergetic 459 MeV antineutrinos within the Sun.  Previous
  analyses omit this possibility as it necessarily involves an
  electroweak suppression factor, and instead search for the
  unsuppressed 20-50 MeV neutrinos produced via two-stage proton decays.  By
  accounting for the relative difference in interaction cross section and experimental background at typical
  neutrino detection experiments, we demonstrate that this
  new channel in fact possesses greater discovery potential.
  As a case in point, using 5326 live days of Super-Kamiokande exposure we find that
  $2\;\sigma$ ($3\;\sigma$) deviations in the 20-50 MeV channel
  are amplified to $3\;\sigma$ ($4.6\;\sigma$) deviations in the 459 MeV case. 
  Exploiting correlations between these two channels may also offer even greater statistical power.
\end{abstract}

%\pacs{xxx.xxx}

%\preprint{xxx-xxx}

\maketitle

%\tableofcontents

%
%
%
%
%

\section{Introduction}
\label{sec:intro}

% Magnetic monopoles are a generic feature of BSM theories

Arising naturally from the spontaneous breaking of non-Abelian
symmetries, magnetic monopoles are arguably one of the most plausible
facets of physics beyond the Standard Model.  In particular, if the
hypothesis of a grand unification of physical forces is indeed
correct, then such phenomena are perhaps unavoidable.

% However, thus far they have evaded detection

Of course, despite their theoretical ubiquity magnetic monopoles also
appear to be in short supply in our visible universe.  Several decades
of experimentation have yielded only a series of ever-tightening
constraints \cite{Burdin:2014xma}, albeit with a few tantalising
events which were later ruled out as monopole candidates
\cite{Price:1975zt,Cabrera:1982gz}.  Further efforts are ongoing, in
particular at the MOEDAL experiment at the LHC
\cite{Acharya:2014nyr}. 

Given the relatively inaccessible scale of grand
unification to present-day particle physics, it is notable that remnant magnetic
monopoles may conceivably provide the most accessible experimental signature available to our
low-energy world.
Another primary motivation for these efforts also lies in the success of the inflationary
paradigm \cite{Ade:2015lrj}, which suggests that any observation of superheavy magnetic monopoles is exceedingly unlikely.
If found any topological relics of this nature would pose a very serious problem for inflationary theory, adding weight to their already huge experimental significance.

Furthermore, given their unusual properties it is also expected that
even a single monopole can leave a highly distinctive signature,
aiding any discovery efforts \cite{Pinfold:2009oia}.
Indeed, of the possible monopole search strategies available to
experimentalists, perhaps the most intriguing relies upon one
particularly exotic property they possess.  As originally established
by Callan and Rubakov \cite{Callan:1982au,Rubakov:1982fp}, certain
types of magnetic monopole are able to directly catalyse the decay of
protons into positrons, without relying on superheavy gauge bosons or
other intermediate states.

Since this leads to a cross-section lacking any of the usual
supression factors, it is expected that these processes can occur at
the rather rapid rates characteristic of the strong interaction.
Furthermore, given the amount of energy liberated in such an event, it
is then expected that it may have particularly noticeable effects in
environments where we expect magnetic monopoles to accumulate, such as
stellar interiors.

% Existing studies of this channel are incomplete

Naturally, there have been a number of studies oriented around these
phenomena, in neutron stars \cite{Kolb:1982si}, white dwarfs
\cite{Freese:1998es} and indeed our own Sun \cite{Arafune:1983sk}.
Further searches have also been performed looking for nucleon decays
arising from the passage of magnetic monopoles through detector arrays
\cite{Ambrosio:2002qq,Aartsen:2014awd,Aartsen:2015exf}.

However, as we will demonstrate in the following there has been also a
subtle omission in the theoretical underpinning of some of these
efforts, which has yet to be exploited.
%
% As such, we propose to search for high-energy solar antineutrinos
%
More specifically, whilst the resultant neutrinos offer a particularly
useful hallmark of typical proton-decay processes, especially those
occurring inside the Sun, existing searches focus entirely on
two-stage processes such as
$p\to \mu^++K\to e^++\nu_e+\overline \nu_\mu+X$. 
%As we will show in
The resulting neutrino flux is
mostly from $\pi^+$ Decay At Rest (DAR), which carries a
characteristic energy ranging from $0$ to $52.8$ MeV, and peaking at
$\sim 35 \; \mathrm{MeV}$~\cite{Arafune:1983sk}. 
It is also well known that protons cannot directly decay to neutrinos via GUT monopoles carrying only $SU(3)\otimes U(1)$ charge. That
is to say, processes such as $p\to \pi^++\overline\nu_e$ are apparently
forbidden~\cite{Ellis:1982bz,Bais:1982hm}. 

However, we note that at sufficiently short distances $SU(3)\otimes U(1)$ will be resolved to the (continuous) Standard Model group,
$SU(3)\otimes SU(2)\otimes U(1)$.  In this limit, the restrictions on processes like
$p\to \pi^++\overline\nu_e$ are no longer necessarily valid.
Furthermore, even accounting for the electroweak suppression factor involved, 
the highly monoenergetic nature of the
resulting antineutrino could in fact offer better discovery potential
than is available via two-stage processes.
%
% Short summary of our findings in this article
%
Indeed, in this article we demonstrate precisely this via the following.
\begin{enumerate}
\item Grand-unification monopoles can catalyse the direct decay $p\to
  \overline\nu_e+\pi^+$, leading to an electroweak-suppressed
  monoenergetic 459 MeV antineutrino flux originating from the Sun. 
\item Due to the reduced atmospheric neutrino background and increased
  interaction cross section at higher energies, the resulting
  significance of this signal at typical neutrino detection
  experiments can exceed that of the previously-explored low energy
  neutrino flux arising from unsuppressed monopole-induced proton
  decay. 
\end{enumerate}

The rest of this article is organised as follows. We briefly outline
monopole catalysed proton decay processes in Section
\ref{sec:monop-induc-prot}, including the well-explored $p\rightarrow
\mu^+ + K$ process and our proposed $p\rightarrow \bar \nu_e + \pi^+$
process. We also estimate the solar monopole abundance as a
preparation for the calculation of the neutrino flux.
In Section~\ref{sec:neutrino-flux}, we calculate the neutrino flux
due to the two-stage proton decay (low energy) and the direct decay
(459 MeV) signals. We also calculate the atmospheric neutrino flux
based on \cite{Honda:2015fha}, in preparation for background estimation.
In Section~\ref{sec:inter-at-detect}, we take Super-Kamiokande as an example and
compute results for the three detection channels, namely scattering from
electrons, protons, and oxygen nuclei. For the neutrino-nuclei cross
section, we use the Fermi gas model only for the 459 MeV neutrinos,
neglecting the low energy neutrino-oxygen scattering in
line with Ref.~\cite{Ueno:2012md}. We also perform an energy cut to
suppress the atmospheric neutrino background while keeping most of the
signal events. 
In Section~\ref{sec:significance}, we compare the statistical
significance of the two channels, and demonstrate that the new, high-energy channel
possesses better discovery potential.
In closing we briefly summarise our result, and indicate directions for future research. 
 
\section{Monopole-induced proton decay}
\label{sec:monop-induc-prot}

% Monopoles can catalyse exotic processes
\subsection{The Callan-Rubakov effect}

As previously noted, one of the more interesting facets of monopole
physics is the possibility of unsuppressed `exotic' processes
occurring, including those which may violate ordinarily conserved
global symmetries.  In practice there are several mutually compatible
interpretations of how these phenomena can occur.

In one picture, we may imagine the monopole as being surrounded by a
cloud of fermion condensate, polarising the vacuum.  This is
explicitly supported by the computation of matrix elements in a
monopole background, giving
$\langle\overline\psi_L\psi_R\rangle\sim1/r^3$, where $r$ is the
radial distance from the monopole core \cite{Rubakov:1982fp}.  Since
monopoles couple with the inverse of the usual electromagnetic
coupling, strong coupling phenomena of this nature should of course
not be surprising.  Incoming fermions are then able to scatter from
this vacuum polarization, leading to processes $\psi_L+M\to \psi_R+M$.

Alternatively, toy models suggest that the dyon mode of the monopole
may play a crucial role \cite{Polchinski:1984uw}.  Therein, charged
fermions passing through the core of the monopole are
absorbed, exciting the dyon mode.  This excitation is unstable and so
will subsequently decay into lighter charged fermions, allowing global
symmetries to be violated.

A third, and far more heuristic perspective is to note that under
S-duality, monopoles and superheavy gauge bosons are exchanged
\cite{Montonen:1977sn}.  Disregarding the fact that this duality is
likely inapplicable to the real world due to an absence of sufficient
supersymmetry, this suggests that the usual process of proton decay
via $X$ or $Y$ bosons may have a dual description in terms of proton
scattering from a magnetic monopole, leading to the same conclusions.

\begin{figure}[h]
  \includegraphics[width=0.5\linewidth]{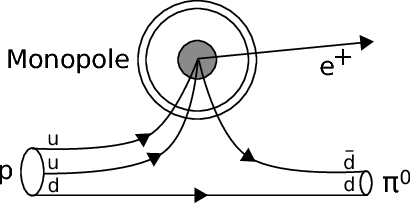}
  \caption{Heuristic diagram of monopole-induced proton decay. The
    monopole provides an effective baryon-number violating
    four-fermion vertex, allowing nucleons to decay.
    Figure reproduced from \cite{Aartsen:2014awd}.}
  \label{fig:monopole-diagram}
\end{figure}

In any case, we can keep in mind the diagrammatic logic of figure
\ref{fig:monopole-diagram}, where the monopole effectively supplies a
baryon-number violating four-fermion vertex.
%
% In the GUT theory context this (usually) leads to proton decay
%
For our purposes we will focus on the monopoles which arise in GUTs
(Grand Unified Theories), and in particular the (equivalent)
minimally-charged monopoles which occur in $SU(5)$ or $SO(10)$
theories.  We note that in some models, such as Pati-Salam~\cite{Chamseddine:2013rta,Aydemir:2015nfa},
%as usual Pati-Salam is not a GUT. Only Pati-Salam from SO(10) or NCG
%is treated as GUT.
there is no monopole catalysis of proton decay, and in others, such as flipped $SU(5)$, there are no monopoles whatsoever \cite{Dawson:1982sc}.

% For SU(3)\otimes U(1) this gives proton decay to neutrinos via two
% stage processes

Once this symmetry is fully broken down to the Standard Model gauge
group, there will be monopoles left over from each symmetry breaking
phase transition.  We can specify each monopole up to gauge
equivalence via the orientation it possesses within $SU(5)$, or more
specifically by the embedding of the $SU(2)$ subgroup it defines.
  For $SU(3)\otimes SU(2)\otimes U(1)$ there are two choices
for minimal monopoles, up to colour equivalence, given by the
diag$(0,0,1,-1,0)$ and diag$(0,0,1,0,-1)$ embeddings
\cite{Liu:1996ea}.

As we expect monopoles such as these accumulate within highly massive,
long-lived objects such as our own Sun, the resultant high-energy
neutrino flux from proton decay then offers a unique channel to infer
their presence, and thereby test the grand-unification
hypothesis.

\subsection{Monopole-sourced two-stage solar neutrino production}
\label{sec:monop-sourc-solar}

In Refs.~\cite{Dokos:1979vu,Lazarides:1980va,Bais:1982hm}, it is shown
that monopoles that carry magnetic $U(1)$ charge $2\pi \;
\mathrm{diag}(1/3, 1/3, 1/3, -1, 0)$ and
magnetic $SU(3)$ charge $2\pi\; \mathrm{diag} (- 1/3, -
1/3, 2/3, 0,0) $ can exist in the embedding of $SU(3)
\otimes U(1) \subset SU(5)$. The magnetic $SU(5)$ charge before
and after electroweak symmetry breaking is then $2\pi \; \mathrm{diag} ( 0, 0, 1, -1,0)$. 
It is straightforward to see that these
 monopoles
carrying strictly $SU(3)\otimes U(1)$ charge cannot offer direct decay
modes to a neutrino, since neutrinos have no electromagnetic or colour charge
and are as such decoupled.
To find the selection rules for processes allowed by these monopoles,
we must construct the fermion doublets selected from the GUT
multiplets via the `monopole' $SU(2)$ subgroup.
As demonstrated in \cite{Bais:1982hm}, neglecting phases these are
\begin{align}
  \begin{pmatrix}
    \overline u_b\\
    u_r
  \end{pmatrix}_L\,,\quad
  % \begin{pmatrix}
  %   \overline u_r\\
  %   u_b
  %	\end{pmatrix}_L\quad
  \begin{pmatrix}
    d_g\\
    e^+
  \end{pmatrix}_L\,,
  % \quad
  % \begin{pmatrix}
  %   e^-\\
  %   \overline d_g
  %	\end{pmatrix}_L\,,
\end{align}
and their conjugates.  In line with the comments of the previous
section, it is straightforward to recognise these as corresponding to
the decay modes of an $SU(5)$ $X$ boson.

By contracting these doublets we can then construct the allowed
operators in the monopole background, such as
$(\overline u_r u_b)(\overline e d_g)$+h.c., which still conserve
electric charge and $B-L$ quantum number.  Computing these in a
monopole background, we find no mass suppression factors, nor coupling
constant dependence.  As such, the corresponding proton decay cross
section is expected to be purely geometric, and so dictated largely by
the size of the proton \cite{Rubakov:1982fp}.

% Branching fraction

Following \cite{Bais:1982hm}, we can then use these points to estimate
the branching fraction of proton decays which produce neutrinos.  The
primary decay channel is $p\to e^++\pi^0$, relative to which the
dominant neutrino producing decay is $p\to \mu^++K$ \footnote{The
  $p\to \mu^++\pi^0$ process is forbidden since it only contains a
  single second-generation fermion}.  Since these processes occur at
geometric rates, it was argued therein that
\begin{align}
  \Gamma\left(p\to e^+\pi\right):\Gamma\left(p\to\mu^+K\right)\simeq 1:(m_d/m_s)^2\,,
\end{align}
the overall scale of any individual catalysed process being set
broadly by the Compton wavelengths of the particles involved.
This ratio depends sensitively on the choice of quark masses, in that
if we take their values from short-distance current algebra then
$(m_d/m_s)^2\sim 1/400$, whilst from their constituent masses
$(m_d/m_s)^2\sim 1/2$.  

 As with instantons, anomalous monopole-induced processes are tied to the presence of fermion zero modes.
Indeed, we can identify the analog of the resulting `t Hooft vertex in Fig.~\ref{fig:monopole-diagram}. 
It is the bare mass appearing in the Lagrangian which is relevant for these zero modes, since they are defined at the level of the equations of motion.
As in the instanton case this is then the mass appearing in their
corresponding fermion determinants, and hence controlling the overall
rate of these anomalous processes. Therefore, we use the
short-distance current algebra mass in the rest of this article.
Conversely, constituent masses are generally understood to arise from
the large binding energy associated to the non-perturbative `gluing'
of free quarks into colour-neutral objects. In this context, they can
only be expected to appear at higher orders in perturbation theory. 

As with ordinary GUT proton decay, there is
also a kinematic suppression to be accounted for.  In this instance,
assuming it is unchanged from the monopole-free context, this gives an
additional factor of $\sim 1/2$.
The resultant neutrino signal, peaked around 35 MeV
\cite{Arafune:1983sk}, then forms the basis for the most stringent
present-day constraints on proton-decay induced solar neutrinos, by
virtue of the Super-Kamiokande experiment \cite{Ueno:2012md}.

% For SU(3)\otimes SU(2)\otimes U(1), direct decay is possible
\subsection{Direct $\nu$-producing proton decay modes}

It is firstly notable that in the context of proton decay within GUT theories 
there are a number of possible channels, some of which include direct
decay of proton to a neutrino.  Indeed, as explored in \cite{Sakai:1981pk} there
are GUTs for which the dominant `ordinary' decay mode is
$p\to K^++\nu$, rather than $p\to e^++\pi^0$.  Whilst this may not be
possible directly in the monopole context, due to the relative
rigidity of the Callan-Rubakov formalism, it is nonetheless suggestive
that alternative channels of monopole-induced direct decays to
neutrinos should be explored.

Let us then consider the minimal GUT monopoles lying in the embedding
$SU(3)\otimes SU(2)\otimes U(1) \subset SU(5)$. There exist monopoles
with the following magnetic charge.
\begin{align}
  \label{eq:mono-magnetic-charge}
  U(1) :  \indent g_1 h_{Y}
  & = 2\pi \; \mathrm{diag} \left( \frac{1}{3},
                             \frac{1}{3}, \frac{1}{3}, -\frac{1}{2}, -
                             \frac{1}{2}\right), \cr
    SU(2) : \indent g_2 \sigma_3
  & =
    2\pi \; \mathrm{diag} \left( 0, 0, 0, \frac{1}{2}, -\frac{1}{2}\right),
    \cr
    SU(3) : \indent g_3 \tau_8
  & =
    2\pi \; \mathrm{diag} \left( - \frac{1}{3}, - \frac{1}{3}, \frac{2}{3},
    0,0\right), 
\end{align}
where $g_1, g_2, g_3$ are the coupling constants of each gauge
subgroup,  $h_Y$ is the $U(1)$ hypercharge generator, and $\sigma_3
(\tau_8)$ is the (last) diagonal element of the Pauli (Gell-mann) matrices, respectively. The
$SU(5)$ magnetic charge then reads $2\pi \; \mathrm{diag} (
0,0,1,0,-1)$. We can again construct the doublets
\begin{align}
  \begin{pmatrix}
    e^+\\
    u_g
  \end{pmatrix}_L\,,\quad
  \begin{pmatrix}
    \overline d_r\\
    u_b
  \end{pmatrix}_L\,,\quad
  \begin{pmatrix}
    \overline{\nu}_e\\
    d_g
  \end{pmatrix}_L\,,
\end{align}
and their conjugates, by acting on the $\overline{\textbf{5}}$ and
$\textbf{10}$ multiplets with the `monopole' $SU(2)$ generator.  It is
straightforward to recognize the result as corresponding to the decay
modes of an $SU(5)$ $Y$ boson.

Contracting as before we then find the effective four-fermion operator
$(\overline d_r u_b)(\overline \nu_e d_g)+h.c.$, which allows
$p\to \overline\nu_e+\pi^+$.  Since this is a simple two-body decay,
the resultant neutrino energy should be $\mathcal{O}(500)$ MeV.
%
% Branching fraction
%
As a single stage decay involving only first
generation fermions, it will also be the dominant neutrino production
channel for $SU(3)\otimes SU(2)\otimes U(1)$ monopoles.  Typically it
should carry a 50\% branching fraction, along with $p\to e^++\pi^0$.
The process $p\to \overline\nu_\mu+K^+$ is also possible, but again
with a relative suppression factor of $(m_d/m_s)^2$.

Since this process relies upon the electroweak component of the magnetic field, once electroweak symmetry is broken we can expect a characteristic suppression factor of $(m_p/m_W)^2$ to enter, since the scale of `ordinary' $SU(3)\otimes U(1)$ monopole-induced proton decay is expected to be set by the size of the proton. 
The resulting branching fraction for direct decays to neutrinos then becomes $\sim 10^{-4}$.

Although we can expect monopoles corresponding to both diag$(0,0,1,-1,0)$ and diag$(0,0,1,0,-1)$ embeddings to be produced around the GUT phase transition with roughly equal probability, a natural concern at this point is the stability of these latter relics under the electroweak phase transition.  Indeed, it is suggested in
\cite{Lazarides:1980va} that electroweak strings could form connecting
these monopoles and their antimonopoles, leading to their rapid
coannihilation.  However, as argued in \cite{Gardner:1983uu} it is
also known that $\pi_1(SU(2)\otimes U(1)_Y/U(1))=0$, and so this does
not seem mathematically plausible.

Another possibility is that these monopoles are `converted' somehow
during this phase transition into the $(0,0,1,-1,0)$ monopoles, and
are hence absent at the present day \cite{Liu:1996ea}.  However, there
is no known mechanism to achieve this, and since these monopoles carry
differing fractional electric charges such a process would naively
appear to violate charge conservation.

A third possibility, discussed in \cite{Gardner:1983uu,Vachaspati:1995yp}, is just that the magnetic electroweak charge
carried by these monopoles is screened below the electroweak scale.
This is of course in line with the known screening of magnetic QCD
charge below $\Lambda_{QCD}$.  As then perhaps the most physically
plausible of the three options, we will assume that this effect is in
operation and proceed accordingly.

Since the only overall suppression we then expect is related to the
scale at which the $SU(3)\otimes SU(2)\otimes U(1)$ structure of the
monopole becomes apparent, the relative suppression factor for direct
vs indirect decays to neutrinos should be
$(m_p/m_W)^2(m_s/m_d)^2$. 
Using the short-distance current algebra values for the quark masses, this then suggests the relative rate of direct to indirect decays to neutrinos should be roughly
$1/16$.  As we will see in
the next section, the relative reduction in neutrino backgrounds and increase in interaction cross section at
459 vs 35 MeV can be more than sufficient to overcome this deficit
experimentally.

\subsection{Monopole abundance in the Sun}
\label{sec:monopoles-sun}

To compute the expected high-energy neutrino flux at Earth-based
detectors, we firstly need to account for the astrophysical monopole
abundance.  The primary constraint in that regard for $10^{16}$ GeV
GUT monopoles comes from the Parker bound, which is based on the
survival of galactic magnetic fields \cite{Parker:1970xv}.  This gives
the galactic monopole flux constraint
\begin{align}
  F_M\lesssim
%\phi_M\times 
10^{-16}\mathrm{cm}{}^{-2}\mathrm{s}{}^{-1}\mathrm{sr}{}^{-1}\,.
% \quad
%   \phi_M\lesssim 1\,.
\end{align}
We define the fraction of $F_M$ to the above bound as $\phi_M$. Stricter bounds are available from the catalysis of nucleon decays
inside neutron stars, however these depend somewhat on the unknown
physics of neutron star interiors \cite{Kolb:1982si}.  Furthermore, it
has been suggested that coannihilation processes may in fact reduce
these limits to the level of the Parker bound
anyway~\cite{Kuzmin:1983by}.

% fraction of monopoles captured

From the age and size of the Sun, we can then estimate the total
number of captured monopoles over the solar lifetime.  In general we
need to account for the infalling velocity and corresponding energy
loss inside the Sun to determine the fraction of monopoles
captured. In Ref.~\cite{Frieman:1985dv} it is estimated that 
this `capture fraction' of $10^{16}$ GeV GUT monopoles is
approximately one, and the number of 
monopoles captured in the Sun is approximately
$N_M \sim 10^{25}\phi_M$. 

% fraction of monopoles left after annihilation

On they other hand, there exists the possibility that the total
monopole number may be diminished by 
coannihilation processes. In this case, one estimates the number of
captured monopoles being $N_M\sim
10^{17}(\phi_M)^{1/3}$~\cite{Frieman:1985dv}. 
A considerable source of uncertainty here is
related to the unknown nature of the magnetic field in the stellar
interior, which could separate monopoles and antimonopoles, and thus
prevent annihilation.
Furthermore, if these processes do occur, the timescale associated to
the annihilation process is largely unknown. In principle the monopole-antimonopole
annihilation cross section should be set by the GUT scale, and their
coannihilation may occur at rates which are negligible for our
purposes~\footnote{The details, including formation of intermediate `monopolonium' bound
  states can be found in \cite{Hill:1982iq}.}.
%Therefore, we neglect the coannihilation of monopoles for our analysis. 

In any case, we primarily aim to demonstrate that the 459 MeV monoenergetic
antineutrino flux offers better discovery potential than the indirect decays previously explored in \cite{Ueno:2012md}.
Since coannihilation will affect both direct and indirect processes equally, it is not a relevant consideration for our purposes.
Therefore, instead of a thorough and detailed classification of
monopole physics, we take a phenomenological `bottom-up' approach and neglect the coannihilation of
monopoles for the rest of the analysis, along the same lines as
Ref.~\cite{Arafune:1983sk}. 

\section{Neutrino Flux}
\label{sec:neutrino-flux}

In this section we calculate the neutrino flux for both high and low energy signals, and the background, which mainly consists of atmospheric
neutrinos. In order to estimate the number of events accurately, we
take into account the neutrino oscillation effect in their propagation
from the Sun to the Earth. 
\subsection{Signal Flux}
\label{sec:signal-flux}

\subsubsection{459 MeV Neutrino}
\label{sec:459-mev-neutrino}

The energy production rate from nucleon decay is firstly
\begin{align}
  \frac{d\epsilon}{dt}=N_M\sigma\rho\beta c\,,
\end{align}
where $\rho$ is the average nucleon density, $\beta c$ is the
thermal nucleon velocity at $T\sim 10^7 \;\mathrm{K} \sim \mathcal{O}(\mathrm{keV})$, and $\sigma$ is the
monopole-nucleon cross-section. It can be expressed as $\sigma=\sigma_0F(\beta)/\beta$, where
$\sigma_0$ is a hadronic cross-section, and $F(\beta)$ is a nuclear
form factor \cite{Arafune:1983sk}.  For hydrogen nuclei,
$F(\beta)\simeq 0.17/\beta$.
The neutrino emission rate and flux on Earth is then
\begin{align}
  F_{\overline\nu}\simeq\frac{1}{4\pi R^2}\frac{dN_{\overline\nu}}{dt}\,,
\qquad
  \frac{dN_{\overline\nu}}{dt}=\frac{BR_{p\to\overline\nu}}{m_pc^2}\frac{d\epsilon}{dt}\,,
\end{align}
where $BR_{p\to\overline\nu}$ is the corresponding branching ratio and
$R$ the average Earth-Sun distance.
Since the $\mathcal{O}$(keV) thermal broadening can be largely neglected, we expect a monochromatic antineutrino flux
\begin{align}
  F_{\overline\nu}\sim\, & 1.4\times 10^4
                           \left(\frac{\sigma_0}{0.1 \mathrm{mb}}\right)
                           \left(\frac{\beta}{10^{-3}}\right)
                           \left(\frac{BR_{p\to\overline\nu}}{10^{-4}}\right)
                           \left(\frac{N_M}{10^{25}}\right)\,\mathrm{cm}{}^{-2}\mathrm{s}{}^{-1}\mathrm{sr}{}^{-1}\,,
\end{align}
of characteristic energy
\begin{align}
  E_{\overline{\nu}}=(m_p^2-m_{\pi^+}^2)/2m_p = 458.755\; \mathrm{MeV}\,. 
\end{align}
Since different neutrino flavours interact differently, we need to
account for the oscillation effects in their propagation from the Sun to
the Earth. 
\begin{align}
  \label{eq:oscillation}
  P(\nu_\alpha \rightarrow \nu_\alpha )
  & =
    1-
    4|U_{\alpha 2} |^2 ( 1- |U_{\alpha 2}|^2) \sin ^2
    \frac{\Delta_{21}}{2}
    -4 |U_{\alpha 3} |^2 ( 1- |U_{\alpha 3}|^2 ) \sin^2
    \frac{\Delta_{31}}{2}
    \cr &
          \indent
    + 2|U_{\alpha 2}|^2|U_{ \alpha 3}|^2
        \left ( 4\sin^2 \frac{\Delta_{21}}{2} \sin^2 \frac{\Delta_{31}}{2}
    +\sin \Delta_{21} \sin \Delta_{31} \right )\,,
    \cr
      P(\nu_\alpha \rightarrow \nu_\beta )
  &=
        4|U_{\alpha 2} |^2 |U_{\beta 2} |^2 \sin ^2
    \frac{\Delta_{21}}{2}
        +4|U_{\alpha 3} |^2 |U_{\beta 3} |^2 \sin ^2
    \frac{\Delta_{31}}{2}
    \cr
  &   ~~ + 2 \Re(U_{\alpha 3}^* U_{\beta 3} U_{\alpha 2} U_{\beta 2}^*)
    \left ( 4\sin^2 \frac{\Delta_{21}}{2} \sin^2 \frac{\Delta_{31}}{2}
    +\sin \Delta_{21} \sin \Delta_{31} \right )
    \cr
  & ~~
    + 4 J_{(\alpha,\beta)} \left ( \sigma^2 \frac{\Delta_{21}}{2} \sin
    \Delta_{31} - \sin^2 \frac{\Delta_{31}}{2} \sin \Delta_{21} \right
    ),
\end{align}
where we use the same notation as
Ref.~\cite{Agarwalla:2013tza}, $\Delta_{ij} \equiv \delta
  m_{ij}^2L/2E = 2.534 \left (\frac{\delta
    m_{ij}^2}{\mathrm{eV}^2}\right )
  \left (\frac{\mathrm{GeV}}{E}\right ) \left (
    \frac{L}{\mathrm{km}}\right )$, $J_{(\alpha,\beta)}$ is the
  Jarlskog invariant, $J_{(\alpha,\beta)} = \Im (U_{\alpha 1}^*
  U_{\beta 1} U_{\alpha 2} U^*_{\beta 2}$).
Neglecting matter effects between the Earth and the Sun and using the
average Earth-Sun distance $1.496\times 10^8 \; \mathrm{km}$, the $\bar
\nu_\mu \rightarrow \bar \nu_e$ appearance and $\bar
\nu_e \rightarrow \bar \nu_e$ survival probability is
\begin{align}
  P(\bar
\nu_e \rightarrow \bar \nu_e)
  &  \approx 0.38\,, 
    \cr
    P(\bar
\nu_\mu \rightarrow \bar \nu_e)
  &  \approx 0.47\,, 
\end{align}
where we take $\delta_{CP} = 0$. This gives us
\begin{align}
  F_{\overline\nu_e}\approx\, & 5.3\times 10^3
                           \left(\frac{\sigma_0}{0.1 \mathrm{mb}}\right)
                           \left(\frac{\beta}{10^{-3}}\right)
                             \left(\frac{BR_{p\to\overline\nu}}{10^{-4}}\right)
                           \left(\frac{N_M}{10^{25}}\right)\,\mathrm{cm}{}^{-2}\mathrm{s}{}^{-1}\mathrm{sr}{}^{-1}\,,
                           \cr
  F_{\overline\nu_\mu}\approx\, & 6.6\times 10^3
                           \left(\frac{\sigma_0}{0.1 \mathrm{mb}}\right)
                           \left(\frac{\beta}{10^{-3}}\right)
                               \left(\frac{BR_{p\to\overline\nu}}{10^{-4}}\right)
                           \left(\frac{N_M}{10^{25}}\right)\,\mathrm{cm}{}^{-2}\mathrm{s}{}^{-1}\mathrm{sr}{}^{-1}\,,
\end{align}

\subsubsection{Low Energy Neutrinos}
\label{sec:low-energy-neutrino}

\begin{figure}[t]
  \centering
  \includegraphics[width=.45\textwidth]{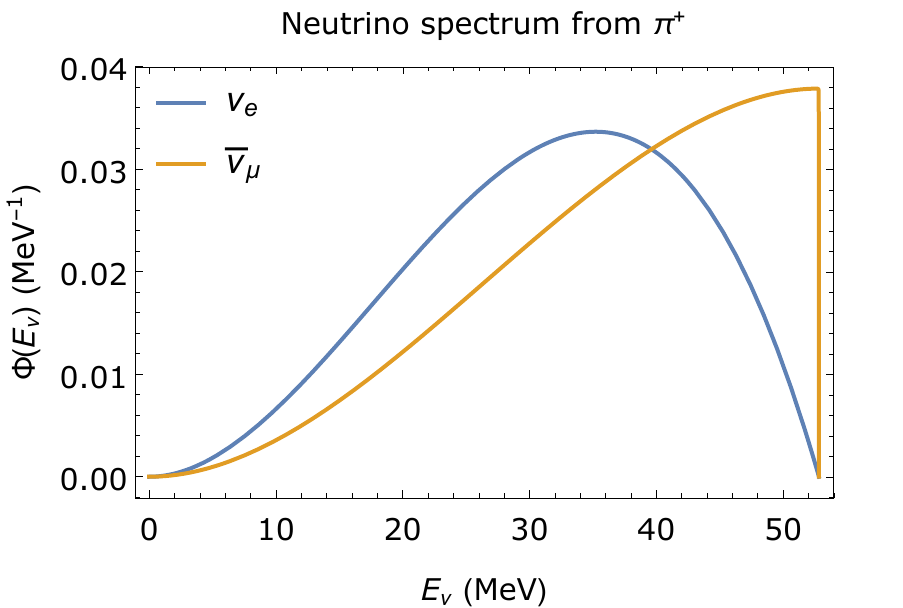}
  \includegraphics[width=.45\textwidth]{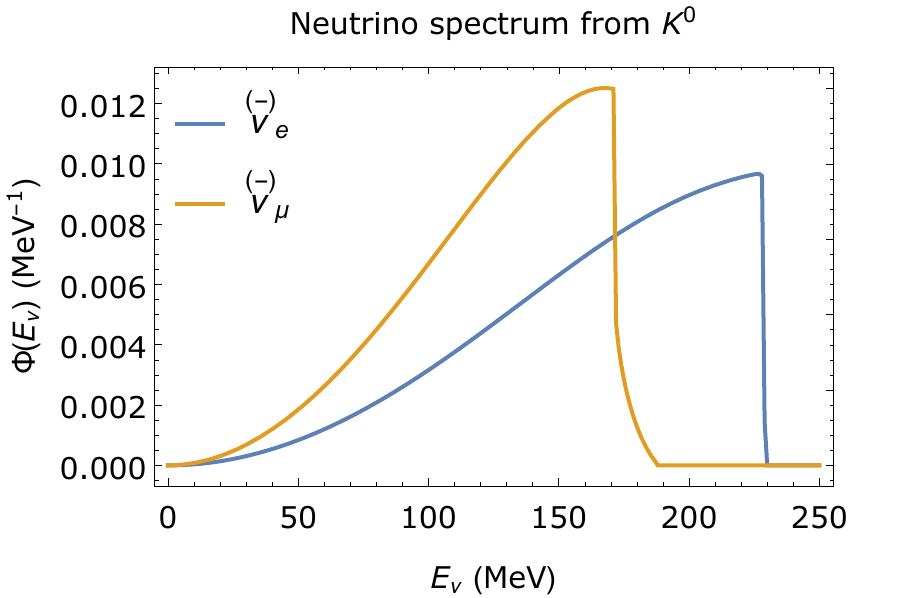}
  \caption{Neutrino spectrum from $\pi^+$ decay. }
  \label{fig:pion-kaon-spectrum}
\end{figure}

Now let us estimate the low energy neutrino flux.
At low energy, the neutrino source is from the kaon and muon produced
in $p\rightarrow K^0 + \mu^+$. 
The relevant decay chains of $K^0 (K_S, K_L)$  are 
\begin{align}
  \label{eq:kaon-decay}
  % K_S & \rightarrow \pi^0 \pi^0 (30.69\%), \cr
          K_S & \rightarrow \pi^+ \pi^-\,,  &(69.20\%) \cr
  % &\indent  \pi^+  \rightarrow \mu^+ +  \nu_\mu,  & (99.99\%) \cr
  % &\indent         \mu^+ \rightarrow e^+ + \bar \nu_\mu + \bar
  %   \nu_e, &  (\approx 100\%) \cr
             %
    K_L & \rightarrow \pi^\pm e^\mp \overset{(-)}{\nu_e}\,,
          &  (40.55\%) \cr
  % &\indent  \pi^+  \rightarrow \mu^+ +  \nu_\mu, &(99.99\%) \cr
  % &\indent         \mu^+ \rightarrow e^+ + \bar \nu_\mu + \bar
  %   \nu_e, &  (\approx 100\%) \cr
             %
             K_L & \rightarrow
                   \pi^\pm \mu^\mp \overset{(-)}{ \;\nu_\mu}\,, & (27.04 \%) \cr
                                              K_L & \rightarrow \pi^+
                                                    \pi^- \pi^0\,. &
                                                                   (12.54\%)
\end{align}
The $\pi^-$ are immediately absorbed by the nuclei in the Sun while
the $\pi^+$ and $\mu^+$ go through the following decay process.
\begin{align}
  \label{eq:pion-source}
  &\indent  \pi^+  \rightarrow \mu^+ +  \nu_\mu\,,  & (99.99\%) \cr
  &\indent         \mu^+ \rightarrow e^+ + \bar \nu_\mu + \bar
    \nu_e\,. &  (\approx 100\%) 
  % \pi^+ & \rightarrow \mu^+ +  \nu_\mu, \cr
  % \mu^+ &\rightarrow e^+ + \bar \nu_\mu + \bar \nu_e.
\end{align}
There are then three separate fluxes we need to consider; the neutrinos
directly from $K^0$ decay, the neutrinos from $\pi^+$ decay, and the
neutrinos from $\mu^\pm$ decay. The neutrino spectra from kaons and muons are
shown in Fig.~\ref{fig:pion-kaon-spectrum}, while the neutrinos from $\pi^+$ are
monoenergetic at $29.8$~MeV. The normalisation for these neutrinos are listed as
follows.
\begin{align}
  \label{eq:neutrino-normalization}
 F_{\bar \nu_e}^{K} = 
 F_{\nu_e}^{K} 
  % &  =
  %                     50\% \times 40.55\% \times 1/2 \times
  %   16 \times
  %   1.4 \times 10^4 \; \mathrm{cm}^{-2} \mathrm{s}^{-1}
  %                     \mathrm{sr}^{-1}, \cr
  & =
    1.62   \times 1.4 \times 10^4 \; \mathrm{cm}^{-2} \mathrm{s}^{-1}
                      \mathrm{sr}^{-1}, \cr
 F_{\bar \nu_\mu}^{K} = 
 F_{\nu_\mu}^{K} 
%  &  =
    %                   50\% \times 27.04\% \times 1/2 \times
    % 16 \times
    % 1.4 \times 10^4 \; \mathrm{cm}^{-2} \mathrm{s}^{-1}
    %                   \mathrm{sr}^{-1}, \cr
  & =
    1.08   \times  1.4 \times 10^4 \; \mathrm{cm}^{-2} \mathrm{s}^{-1}
    \mathrm{sr}^{-1}, \cr
    F_{\bar \nu_e}^{\mu^-}=
    F_{ \nu_\mu}^{\mu^-}
  % & =
  %                     50\% \times 27.04\% \times 1/2 \times
  %   16 \times
  %   1.4 \times 10^4 \; \mathrm{cm}^{-2} \mathrm{s}^{-1}
  %                     \mathrm{sr}^{-1}, \cr
  & =
    1.08   \times  1.4 \times 10^4 \; \mathrm{cm}^{-2} \mathrm{s}^{-1}
    \mathrm{sr}^{-1}, \cr
    F_{\nu_\mu}^{\pi^+}
  % &= \left (
  %   50\% \times 69.20\% + 50\% \times 40.55 \% \times 1/2 + 50 \%
  %   \times27.04 \% \times 1/2 + 50 \% \times 12.54 \%
  %   \right )\cr
  % &\indent
  %   \times 16 \times
  %   1.4 \times 10^4 \; \mathrm{cm}^{-2} \mathrm{s}^{-1}
  %   \mathrm{sr}^{-1}, \cr
  & =
    9.24 \times
    1.4 \times 10^4 \; \mathrm{cm}^{-2} \mathrm{s}^{-1}
    \mathrm{sr}^{-1}, \cr
    F_{\bar \nu_\mu}^{\mu^+}
    =
    F_{ \nu_e}^{\mu^+}
  % &= \left (
  %   50\% \times 69.20\% + 50\% \times 40.55 \% \times 1/2 + 50 \%
  %   \times27.04 \% \times 1/2 + 50 \% \times 12.54 \%
  %   +50\% \times 27.04\%
  %   \right )\cr
  % &\indent
  %   \times 16 \times
  %   1.4 \times 10^4 \; \mathrm{cm}^{-2} \mathrm{s}^{-1}
  %   \mathrm{sr}^{-1}, \cr
  & =
    11.41  \times
    1.4 \times 10^4 \; \mathrm{cm}^{-2} \mathrm{s}^{-1}
    \mathrm{sr}^{-1},
\end{align}
where we saturate the Parker bound, as the neutrino flux scales
linearly with the monopole flux. 
It is observed that the kaon neutrino spectrum ranges from zero to
$\sim$ 200 MeV,
and peaks at high energy. This means that below 50 MeV the effect is
negligible (about 1/10 compared to the pion/muon neutrinos), and the flux is
not energetic enough to swamp our 459 MeV neutrino signal. Therefore it
doesn't affect either our high or low energy neutrino analysis, and so the low energy neutrino signal mostly consists of
$\pi^+$ DAR flux. 
Next we take into account the neutrino oscillation probability. 
\begin{align}
  P(\bar
\nu_\mu \rightarrow \bar \nu_e)
& =
\frac{1}{E_{\nu,max}} \int_0^{E_{\nu, max}} P_{e\mu}(E_\nu)
                                  \Phi(E_\nu) \; d E_\nu\,,
                                  \cr
  & =
\frac{1}{E_{\nu,max}} \int_0^{E_{\nu, max}} P_{e\mu}(E_\nu)
    \; d E_\nu
    \times
    \int_0^{E_{\nu, max}} \Phi(E_\nu) \; d E_\nu    
  \approx 0.27\,, 
\cr
  P(\bar
\nu_e \rightarrow \bar \nu_e)
  &  \approx 0.55\,, \cr
  P(
\nu_\mu \rightarrow \nu_\mu)
    & \approx 0.36\,,
\end{align}
where $\Phi(E_\nu)$ is the neutrino flux normalised to one,
and the decomposition is valid because at this energy scale the
neutrinos are highly oscillatory. 
This yields
\begin{align}
  F_{\overline\nu_e}\approx\, & 5.1\times 10^4
                           \left(\frac{\sigma_0}{0.1 \mathrm{mb}}\right)
                           \left(\frac{\beta}{10^{-3}}\right)
                         \left(\frac{BR_{p\to\overline\nu}}{10^{-4}}\right)
                           \left(\frac{N_M}{10^{25}}\right)\,\mathrm{cm}{}^{-2}\mathrm{s}{}^{-1}\mathrm{sr}{}^{-1}\,,
                           \cr
  F_{\nu_e}\approx\, & 9.2\times 10^4
                           \left(\frac{\sigma_0}{0.1 \mathrm{mb}}\right)
                           \left(\frac{\beta}{10^{-3}}\right)
                         \left(\frac{BR_{p\to\overline\nu}}{10^{-4}}\right)
                           \left(\frac{N_M}{10^{25}}\right)\,\mathrm{cm}{}^{-2}\mathrm{s}{}^{-1}\mathrm{sr}{}^{-1}\,,
                           \cr
  F_{\overline\nu_\mu}\approx\, & 6.2\times 10^4
                           \left(\frac{\sigma_0}{0.1 \mathrm{mb}}\right)
                           \left(\frac{\beta}{10^{-3}}\right)
                         \left(\frac{BR_{p\to\overline\nu}}{10^{-4}}\right)
                           \left(\frac{N_M}{10^{25}}\right)\,\mathrm{cm}{}^{-2}\mathrm{s}{}^{-1}\mathrm{sr}{}^{-1}\,,
                           \cr
  F_{\nu_\mu}\approx\, & 4.9\times 10^4
                           \left(\frac{\sigma_0}{0.1 \mathrm{mb}}\right)
                           \left(\frac{\beta}{10^{-3}}\right)
                         \left(\frac{BR_{p\to\overline\nu}}{10^{-4}}\right)
                           \left(\frac{N_M}{10^{25}}\right)\,\mathrm{cm}{}^{-2}\mathrm{s}{}^{-1}\mathrm{sr}{}^{-1}\,.
\end{align}
%{\color{red}add other neutrino flavors and and redo estimate}

\subsection{Background Flux}
\label{sec:background-flux}
\subsubsection{Solar Neutrino Flux}
\label{sec:constr-from-solar}
The solar neutrino flux around the Earth is estimated
\cite{Robertson:2012ib,Antonelli:2012qu} to be about
$\Phi_{\nu\odot}\sim 6.5 \times 10^{11} \;
\mathrm{cm}^{-2}\mathrm{s}^{-1}$. The energy range of such neutrinos
is predicted by the Standard Solar Model (SSM) to be
$E_{\nu\odot} \lesssim 10\; \mathrm{MeV}$, which is in a very
different energy range to that under consideration. 
As such, the solar neutrino background is for our purposes negligible.

\subsubsection{Atmospheric Neutrino Flux}
\label{sec:constr-from-atmosph}
The atmospheric neutrino flux can be modeled theoretically, such as in
FLUKA~\cite{battistoni2005atmospheric,Battistoni:2002ew}, Bartol~\cite{Barr:2004br}, and HKKM~\cite{Honda:2006qj,Honda:2011nf,Honda:2015fha}.
It is shown in Ref.~\cite{Richard:2015aua} that all three models are consistent with the Super-Kamiokande measurement.
Therefore, we extract the value in Refs.~\cite{Honda:2015fha} to
estimate the atmospheric neutrino background
\begin{align}
  \Phi_{\bar \nu_{e} + \nu_e} &\approx 2.8 \times 10^{-3} \left ( \frac{R_E}{10 \%}\right )\; \mathrm{cm}^{-2} \mathrm{s}^{-1} \mathrm{sr}^{-1}\,, 
  \cr
  \Phi_{\bar \nu_{\mu}+ \nu_\mu} & \approx 6.2 \times 10^{-3} \left ( \frac{R_E}{10 \%}\right ) \; \mathrm{cm}^{-2} \mathrm{s}^{-1} \mathrm{sr}^{-1}\,, 
\end{align}
where $R_E$ is the reconstructed energy resolution of the incident
neutrino at the detector, which determines the bin width. 
For Super-Kamiokande \cite{Richard:2015aua}, the neutrino bin width is
smaller at low energy (sub-GeV) and bigger at high energy
(multi-GeV). 
For the angular resolution, the Sun's angular span is about $6.8\times
10^{-5} \; \mathrm{sr}$, which is smaller than the angular resolution
of most detectors, and so improved directional information could be useful for suppressing the atmospheric neutrino background even
further.

\begin{figure}[htp]
  \centering
  \includegraphics[width=.5\textwidth]{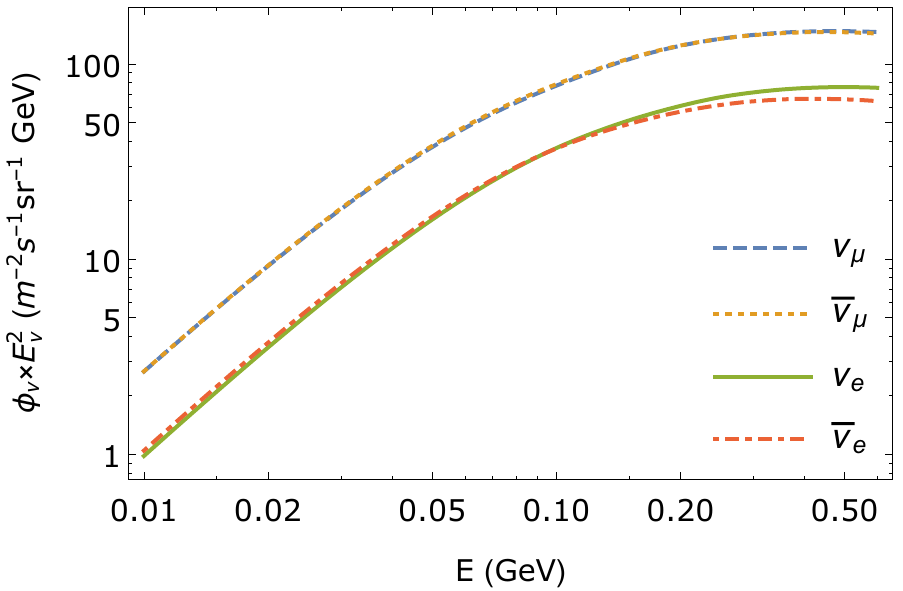}
  \caption{Atmosphere neutrino flux modeled in
    \cite{Honda:2015fha}. The blue (dashed), orange 
    (dotted), green (solid), red (dash-dotted) correspond to
    $\nu_\mu$, $\bar \nu_\mu$, $\nu_e$, $\bar \nu_e$ respectively. 
  \label{fig:atm-nu-flux}}
\end{figure}

\section{Detection Cross Section}
\label{sec:inter-at-detect}
In this section, we take into account the spectrum of the neutrino
flux, and use Super-Kamiokande as a benchmark to calculate the cross
section for a water Cherenkov detector.
In a setup similar to Super-Kamiokande the target particles are electrons,
protons, and oxygen nuclei. Neutrinos interact with them via both neutral current (NC)
and charged current (CC) interactions. In
Ref.~\cite{Ueno:2012md}, only interactions with electrons and protons
are used for the low energy neutrinos, and so following their analysis we then suppose the charged current channel below
$50\;\mathrm{MeV}$ is not clean enough and thus cannot be used for
detection. In addition, we assume that these
interactions can be excluded experimentally, and so do not provide an additional source
of background either. We discuss the three interaction channels one by
one. 

\subsection{Interactions with Electrons}
\label{sec:inter-with-electr}
Even though this is the cleanest channel, as we will see next, the
cross section is also the smallest of the three channels. The
relevant interactions are elastic scattering (ES) processes, as the
inverse muon decay channel has a threshold energy of
$10.92\,\mathrm{GeV}\;$. 
In the Standard Model, the interaction between neutrino flavour $\alpha$ ($\alpha=e,\mu,\tau$) and the electron 
is described at low energies by the effective four fermion interaction
%
%%%%%%%%%%%%%%%%%%%%%%%%%%%%%%%%%%%%%
\begin{equation}
\mathcal{L}_\mathrm{SM} 
\;=\;
-2\sqrt{2}\,G_{F} (\bar{\nu}_{\alpha} \gamma^{\mu} P_L \nu_{\alpha})
\Bigl[\,
g_{\alpha L} (\bar{e} \gamma_{\mu} P_{L} e) + 
g_{\alpha R} (\bar{e} \gamma_{\mu} P_{R} e)
\,\Bigr]\,.
\label{eqn1}
\end{equation}
%%%%%%%%%%%%%%%%%%%%%%%%%%%%%%%%%%%%%
%
The coupling constants at tree level are given by $g_{\alpha R} = \sin^2 \theta_{W}$ and  
$g_{\alpha L} = \sin^2\theta_{W} \pm \frac{1}{2}$, where the lower sign applies for 
$\alpha=\mu$ and $\tau$ (from $Z$ exchange only) and the upper sign applies for $\alpha=e$ (from both $Z$ and $W$ exchange).
For antineutrinos, the values of $g_{\alpha L}$ and
$g_{\alpha R}$ will be reversed.
The differential cross section for neutrino-electron elastic scattering (ES) due to this interaction 
is given by
\begin{equation}
\frac{d{\sigma}_{\nu_{\alpha}}(E_{\nu_\alpha},T_e)}{dT_e} = \frac{2G^2_{F} m_{e}}{\pi} 
\left[
{g}^2_{\alpha L} + {g}^2_{\alpha R}
\left(
1 - \dfrac{T_e}{E_{\nu_\alpha}}
\right)^2 - {g}_{\alpha L} {g}_{\alpha R} \dfrac{m_e T_e}{E_{\nu_\alpha}^2}
\right]
\,,
\label{eqn3}
\end{equation}
where, $m_e$ is the electron mass, $E_{\nu_\alpha}$ is the initial energy of neutrino flavour $\alpha$,
and $T_e$ is the kinetic energy of the recoil electron, which has the range 
\begin{equation}
0 \;\le\; T_e \;\le\; T_{\max}(E_{\nu_\alpha}) 
\;=\; \dfrac{E_{\nu_\alpha}}{1+m_e/2E_{\nu_\alpha}}\,.
\label{Trange}
\end{equation}

\subsubsection{Low Energy Signal}
\label{sec:low-energy-signal}

\begin{figure}[t]
  \centering
  \includegraphics[width=.45\textwidth]{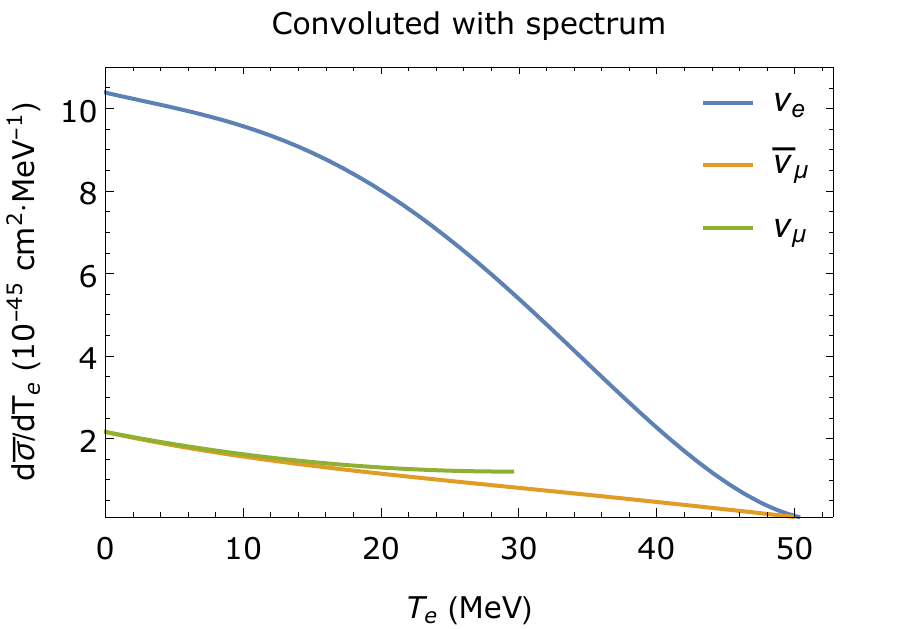}
    \includegraphics[width=.45\textwidth]{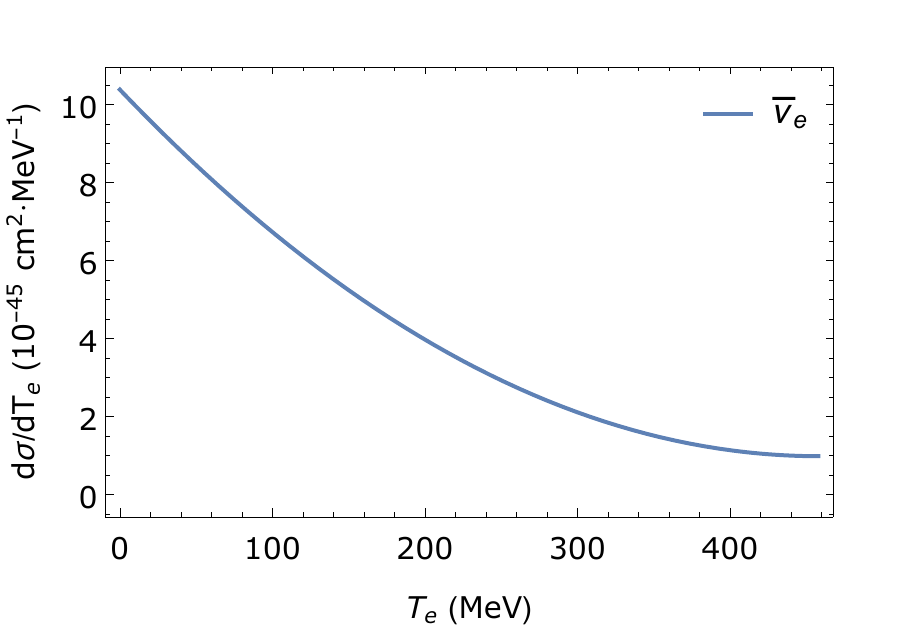}
  \caption{Left: convolution of neutrino spectrum  from $\pi^+$ decay with differential cross section $d\sigma/d T_e$, where $T_e$ is the kinetic
    energy of the recoiling electrons. Right: the electron scattering cross section for $459\;\mathrm{MeV}$ antineutrinos.}
  \label{fig:pion-spectrum}
\end{figure}
We convolute the differential cross section $d\sigma/d T_e$ with the
above spectrum, where $T_e$ is the electron's recoil energy and
\begin{align}
  \label{eq:lowE-electron-convolution}
  \frac{d \tilde \sigma}{d T_e}
  & =
    \int_{E_\nu, min}^{E_\nu, max}
      \frac{d \sigma}{d T_e}(E_\nu)
    \Phi(E_\nu) d E_\nu\,,
\end{align}
as shown in Fig.~\ref{fig:pion-spectrum}.
Following Ref.~\cite{Ueno:2012md}, an energy cut with respect to
the recoiling electron ($T_e> 20$ MeV) is performed to suppress the atmospheric
neutrino background. The total cross section before and after the cut is
\begin{align}
  \label{eq:total-xsec-electron-low-E}
  \sigma_{\nu_e}^{\mathrm{no\;cut}} & =
                   3.05\times 10^{-43} \; \mathrm{cm^2}\,, \cr
  \sigma_{\bar \nu_\mu}^{\mathrm{no\; cut}} & =
                   5.08\times 10^{-44} \; \mathrm{cm^2}\,, \cr
  \sigma_{\nu_e}^{\mathrm{no\; cut}} & =
                   4.48 \times 10^{-44} \; \mathrm{cm^2}\,,
                   \cr
  \sigma_{\nu_e} & =
                   1.16\times 10^{-43} \; \mathrm{cm^2}\,, \cr
  \sigma_{\bar \nu_\mu} & =
                   1.90\times 10^{-44} \; \mathrm{cm^2}\,, \cr
  \sigma_{\nu_e} & =
                   1.17 \times 10^{-44} \; \mathrm{cm^2}\,.                   
\end{align}
Using the 5326 live days of Super-Kamiokande data \cite{Abe:2017aap}, we estimate
the number of events for each flavour to be, after the energy cut
\begin{align}
  \label{eq:number-of-events-low-E-signal}
  N_{\nu_e} \approx 4\,, \quad N_{\bar \nu_\mu} \approx 1\,, \quad N_{\nu_\mu} \approx 0\,.
\end{align}
\subsubsection{High Energy Signal}
\label{sec:high-energy-events}
The differential cross section at high energy is shown in Fig.~\ref{fig:pion-spectrum}.
The total cross section before and after the $20$ MeV cut is
\begin{align}
  \label{eq:nu-e-xsec-highe-total}
    \sigma_{\bar \nu_e}^{\mathrm{no\;cut}} & =
                   1.89\times 10^{-42} \; \mathrm{cm^2}\,, \cr
    \sigma_{\bar \nu_e} & =
                   1.69\times 10^{-42} \; \mathrm{cm^2}\,. 
\end{align}
The number of events is
\begin{align}
  N_{\bar \nu e} & \approx 5\,.
\end{align}
It is observed, due to the small cross section, that even when we saturate the
Parker bound the number of events from electron interaction is very
low. We will next calculate the signal and background from scattering events
with protons and oxygen nuclei.

\subsection{Interactions with Protons}
\label{sec:inter-with-prot}

Neither the low energy neutrinos from pion decay or the 459 MeV
neutrinos are energetic enough to cause Cherenkov radiation of the
recoil protons. 
Therefore, the relevant channel is the inverse beta (muon)
decay process, where
\begin{align}
  \label{eq:neutrinos-proton-int}
  & \bar \nu_e + p \rightarrow e^+ + n\,,\cr
  & \bar \nu_\mu + p \rightarrow \mu^+ + n\,,
\end{align}
and the neutrons emit a 2.2 MeV $\gamma$ when they
combine with protons later. This can be distinguished from the
electron scattering events % used for detection purpose
since neutron tagging became possible after the 2008 Super-Kamiokande IV
upgrade \cite{Zhang:2013tua}.  However, we do not base our analysis on
this new neutron tagging technology. 

\begin{figure}[ht]
  \centering
  \includegraphics[width=.5\textwidth]{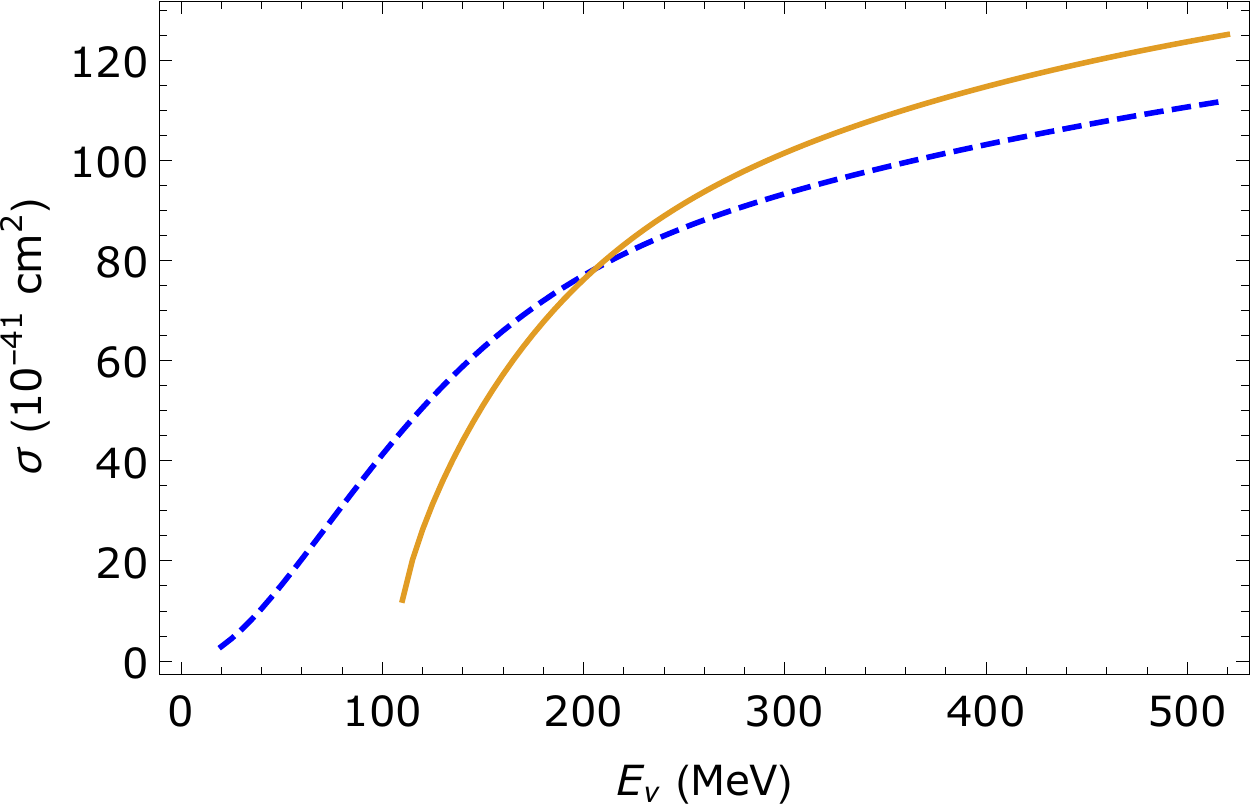}
  \caption{Total cross section of $\bar \nu_e + p \rightarrow  e^+ +
    n$ (blue dashed) and $\bar \nu_\mu + p \rightarrow \mu^+ +
    n$ (orange solid), using the findings of Ref.~\cite{Strumia:2003zx}.}
  \label{fig:xsec-nu-nucleon}
\end{figure}

The inverse muon decay process threshold is
\begin{align}
  \label{eq:IMD}
  E_{\bar \nu_e, thres}
  & =
    \frac{(E_n + E_e)^2 - m_p^2}{2 m_p}
    =
    1.8 \;\mathrm{MeV}\,, \cr
  E_{\bar \nu_\mu, thres}
  & =
    \frac{(E_n + E_\mu)^2 - m_p^2}{2 m_p}
    =
    113.1 \;\mathrm{MeV}\,.
\end{align}
Therefore, inverse muon decay is only relevant for the 459 MeV
neutrinos and absent for low energy neutrinos.  The cross section can be calculated
numerically, or approximated analytically \cite{Strumia:2003zx}. We
follow Ref.~\cite{Strumia:2003zx} and use the analytic expression for
this analysis. The total cross section is reproduced and
shown in Fig.~\ref{fig:xsec-nu-nucleon}. 
\subsubsection{Low Energy Events}
\label{sec:low-energy}

\begin{figure}[htp]
  \centering
  \includegraphics[width=.45\textwidth]{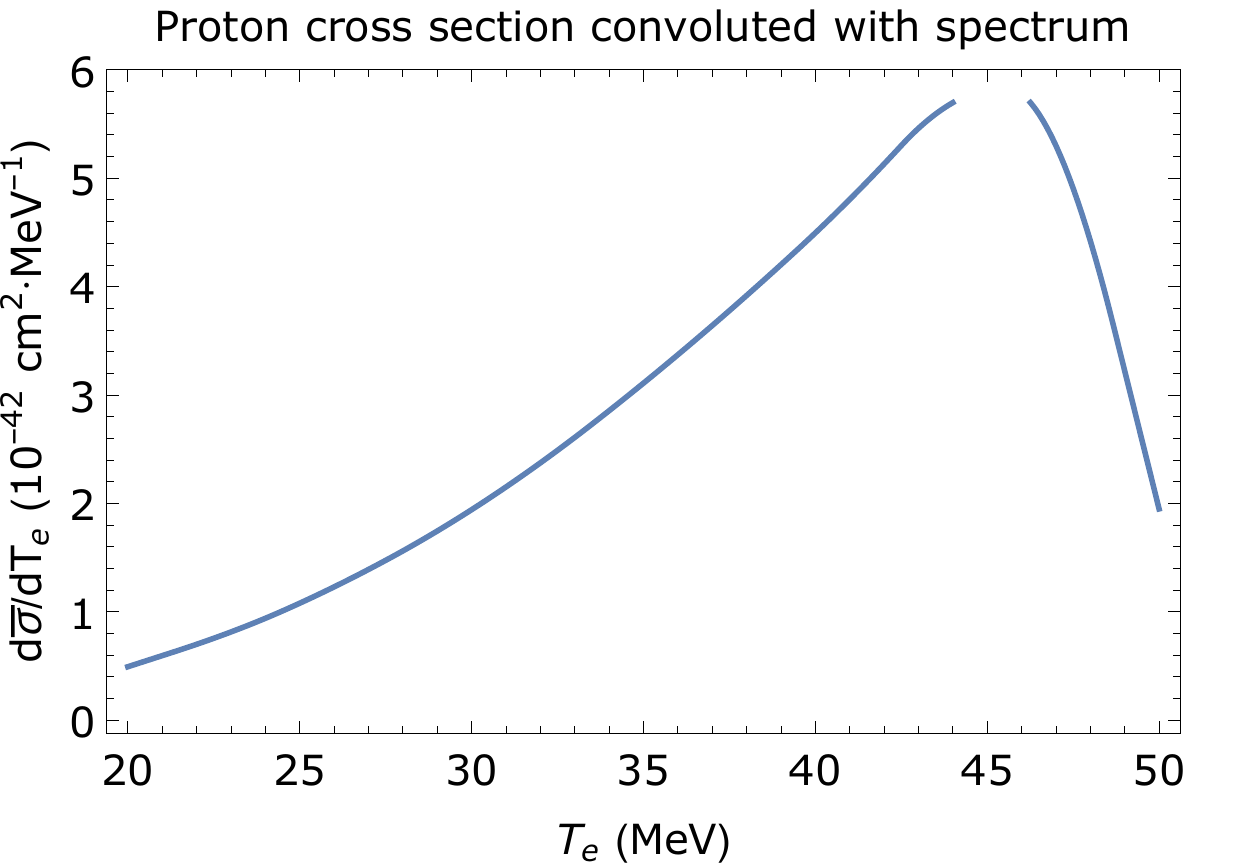}
  \includegraphics[width=.47\textwidth]{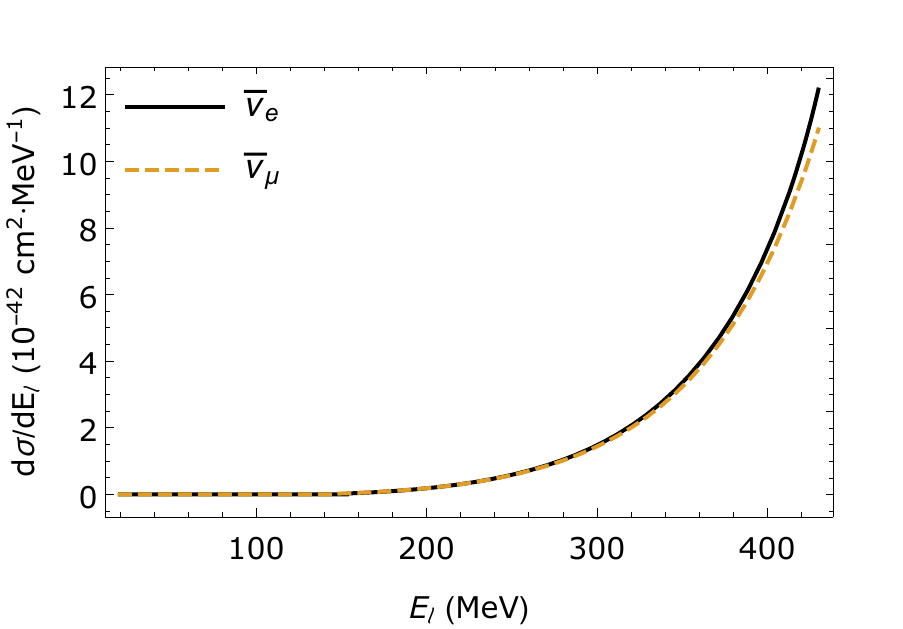}
  \caption{Left: convoluted cross section of low energy $\bar \nu_e p$
    scattering. Right:     Cross section of high energy antineutrino scattering from a
    proton target. The horizontal axis is the energy of the (electron or muon) recoil
    lepton.}
  \label{fig:nuebar-proton-convoluted}
\end{figure}

Integrating the formula in Ref.~\cite{Strumia:2003zx} with the
neutrino spectrum in Fig.~\ref{fig:pion-spectrum}, we show the convoluted differential cross
section in
the left panel of  Fig.~\ref{fig:nuebar-proton-convoluted}.
After the energy cut, the total cross section is
\begin{align}
  \sigma_{\bar \nu_e}
  & = 9.21 \times 10^{-41} \; \mathrm{cm}^2\,.
\end{align}
It is observed that from $\pi^+$ decay no $\bar \nu_e$ is directly
produced. As $\bar \nu_\mu$ propagate to Earth, they oscillate to
$\bar \nu_e$, as detailed in Section \ref{sec:signal-flux}. 
Using the 5326 live days of Super-Kamiokande data \cite{Abe:2017aap}, we estimate
the number of events for $\bar \nu_e$ to be
\begin{align}
  N_{\bar\nu_e}
  \approx 194\,.
\end{align}

\subsubsection{High Energy Events}
\label{sec:high-energy-events}
The differential cross section of 459 MeV antineutrino scattering is shown in
the right panel of  Fig.~\ref{fig:nuebar-proton-convoluted}.  Using the formula from Ref.~\cite{Strumia:2003zx}, the total cross
section is estimated to be
\begin{align}
  \sigma_{\bar \nu_e p}  = 1.08 \times 10^{-39} \; \mathrm{cm^2}\,,\quad
  \sigma_{\bar \nu_\mu p}  = 1.20 \times 10^{-39} \; \mathrm{cm^2}\,.
\end{align}
The number of events is
\begin{align}
  N_{\bar \nu_e}
   = 235\,, \quad
    %618
                   N_{\bar \nu_\mu}  = 324\,. 
                   % 690. \cr      
\end{align}

\subsubsection{Background Events}
\label{sec:background-events}

\begin{figure}[t]
  \centering
  \includegraphics[width=.45\textwidth]{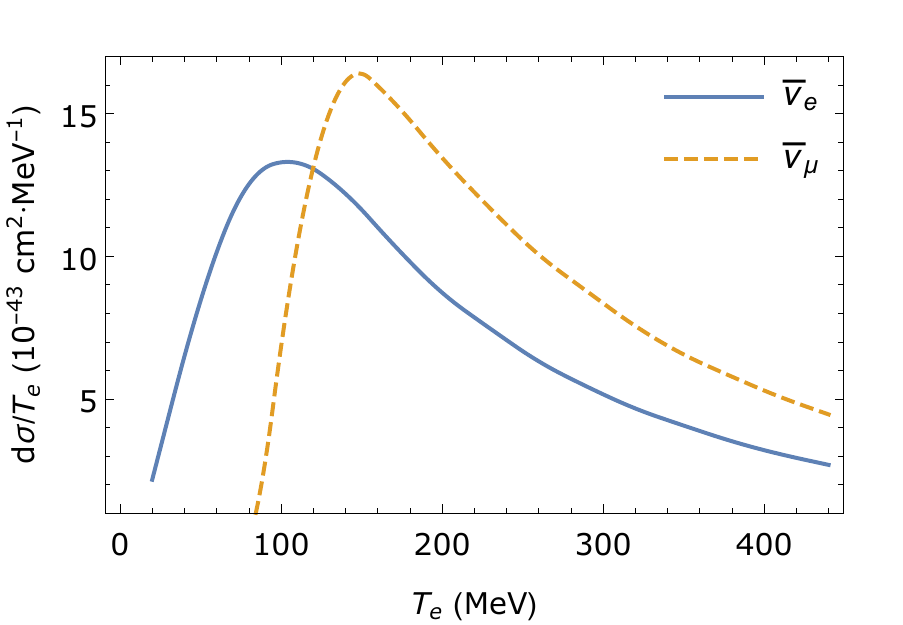}
  \caption{Differential cross section of antineutrino proton scattering,
    convoluted with background flux.}
  \label{fig:xsec-proton-bg}
\end{figure}

The
neutrino/proton differential cross section convoluted with atmospheric
neutrinos is shown in Fig.~\ref{fig:xsec-proton-bg}.
From 20 to 50 MeV, the number of background events is
\begin{align}
  N_{\bar \nu_e} & = 17\,.
\end{align}
At high energy, the number of background events is
\begin{align}
  N_{\bar \nu_e}  =279\,,\quad
  N_{\bar \nu_\mu}  =465\,.
\end{align}

We observe that the signal peaks at high energy, while the
background peaks at below 200 MeV. 
Therefore, we can perform an energy cut to suppress the background. 
With for example a 300 MeV cut, the number of events for signal and background reads
\begin{align}
  N_{\bar \nu_e}^{signal}
   =
    220\,,\quad
                   %578, \cr
    N_{\bar \nu_\mu}^{signal}
   =
    305\,, \quad
                       %649,
  N^{atm}_{\bar \nu_e}  =49\,,\quad
                         N^{atm}_{\bar \nu_\mu}  =122\,, 
  \end{align}
from which we can see the background is suppressed by $\sim  80\%$,
with less than $\sim 10\%$ signal cross section sacrificed. 
In the next section, we will show a more rigorous cut that optimises the
statistical significance.

\subsection{Interactions with Oxygen}
\label{sec:inter-with-oxyg}

As noted in \cite{Ankowski:2012iy,Ankowski:2015lma}, in the energy range of a few
hundred MeV the single nucleon knock-out neutral current
quasi-elastic (NCQE) scattering is significant.
Predictions from various models are compared and shown in
Fig.~1 there, where cross sections for antineutrinos knocking out a single neutron and
a single proton are both around $3\times 10^{-39} \;\mathrm{cm}^2$.
However, as there is no charged lepton emission, we do not use these
processes for detection purposes. 

\begin{figure}[t]
  \centering
  \includegraphics[width=.5 \textwidth]{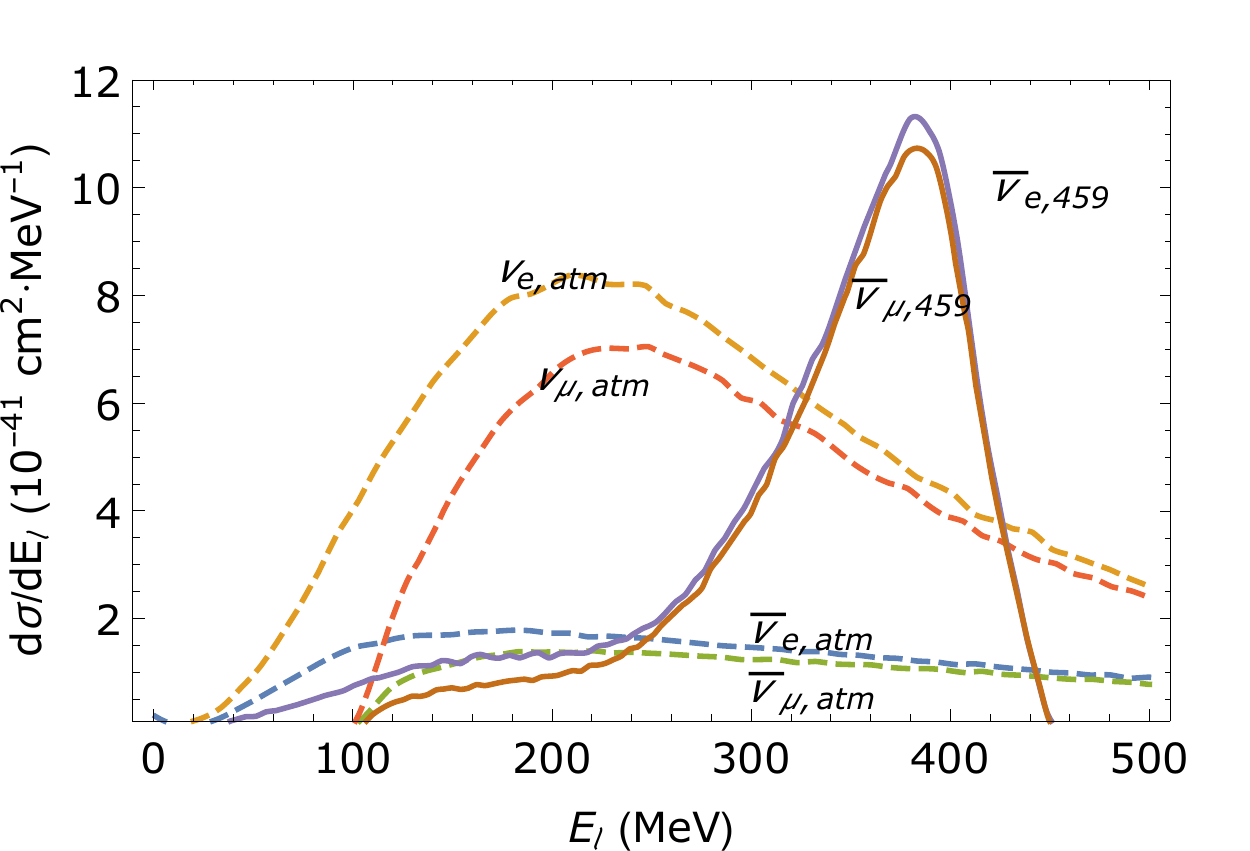}
  \caption{The differential cross section with ${}^{16} O$ of the $459\;\mathrm{MeV}$
    monoenergetic antineutrino signal (solid) and the atmospheric neutrino background
    (dashed). It is observed that the recoiling electrons from atmospheric neutrinos peak at
    around 200 MeV, whilst those from our 459 MeV antineutrinos peak at
    around 400 MeV. }
  \label{fig:O18-xsec}
\end{figure}

Charged current quasi-elastic (CCQE) scattering takes place via the 
following processes
\begin{align}
  \label{eq:CCQE}
  &\bar \nu_e + {}^{16} \mathrm{O} \rightarrow e^+ + {}^{16}\mathrm{N},\cr
  &\bar \nu_\mu + {}^{16} \mathrm{O} \rightarrow \mu^+ +
    {}^{16}\mathrm{N}, \cr
      & \nu_e + {}^{16} \mathrm{O} \rightarrow e^- +
        {}^{16}\mathrm{F},\cr
      & \nu_\mu + {}^{16} \mathrm{O} \rightarrow \mu^- +
        {}^{16}\mathrm{F}.
\end{align}
We use the Monte Carlo software \texttt{NuWro} \cite{Golan:2012wx}
to calculate the $(\nu, {}^{16}O)$ cross section with $459
\;\mathrm{MeV}$ $\bar \nu$ and the atmospheric neutrino flux, shown in Fig.~\ref{fig:O18-xsec}. Again, the cross
section of atmospheric neutrinos is the result of the convolution
\begin{align}
  \label{eq:convolution}
  \frac{d \tilde \sigma}{ d T_\ell} 
  & =
    \int_{E_{\nu, min}}^\infty
    \frac{d \sigma}{d T_\ell}  (E_\nu, T_\ell)
    \Phi(E_\nu)\;
    d E_\nu\,,
\end{align}
where $E_{\nu, min}$ is the minimum energy required by the kinematics
of the process, given a recoil lepton energy $T_\ell$. 
 It is observed
that the convoluted cross section of atmospheric neutrinos peaks at
$\sim 200\;\mathrm{MeV}$. 
Similar to proton scattering, this can be understood as the result of
two competing factors: when the neutrino energy increases, the
neutrino-nucleus cross
section increases, however the atmospheric neutrino flux also decreases, as shown in
Fig.~\ref{fig:atm-nu-flux}. This balance guarantees the
separation of our signal and background.
\subsubsection{High Energy Events}
\label{sec:high-energy-events-3}
The cross section and number of events for 459 MeV antineutrino scattering from oxygen are
\begin{align}
  \sigma_{\bar \nu_e {}^{16} O}
  & =
    1.41\times 10^{-38} \;\mathrm{cm}^2\,, 
    \qquad
    N_{\bar \nu_e {}^{16} O}
    \approx 1538\,,
    \cr
  \sigma_{\bar \nu_\mu {}^{16} O}
  & = 1.26\times 10^{-38} \; \mathrm{cm}^2\,, 
    \qquad
  N_{\bar \nu_\mu {}^{16} O}
    \approx 1699\,.
\end{align}
\subsubsection{Background Events}
\label{sec:background-events-1}
The cross section above 100 MeV and corresponding number of 459 MeV neutrino scattering events with oxygen are
\begin{align}
  \sigma_{\bar \nu_e {}^{16} O}
  & = 2.94 \times 10^{-39} \;\mathrm{cm}^2
    \qquad
    N_{\bar \nu_e {}^{16} O}
   \approx 885\,,
    \cr
  \sigma_{ \nu_e {}^{16} O}
  & = 1.39 \times 10^{-38} \; \mathrm{cm}^2
    \qquad
    N_{ \nu_e {}^{16} O}
   \approx 4188\,,
    \cr
  \sigma_{\bar \nu_\mu {}^{16} O}
  &  = 5.04\times10^{-39}\; \mathrm{cm}^2
    \qquad
  N_{\bar \nu_\mu {}^{16} O}
   \approx 1518\,,
    \cr
  \sigma_{ \nu_\mu {}^{16} O}
  & = 2.20 \times 10^{-38}\;\mathrm{cm}^2
    \qquad
  N_{ \nu_\mu {}^{16} O}
   \approx 6616\,.
\end{align}

Again, if we perform a 300 MeV energy cut, the signal and background
are reduced to
\begin{align}
    N^{signal}_{\bar \nu_e {}^{16} O}
     \approx 1143\,,
    \qquad 
    N^{atm}_{\bar \nu_e {}^{16} O}
&    \approx 330\,,
   \cr
    N^{atm}_{ \nu_e {}^{16} O}
   &    \approx 1457\,,
    \cr
  N^{signal}_{\bar \nu_\mu {}^{16} O}
   \approx 1351\,,
     \qquad
       N^{atm}_{\bar \nu_\mu {}^{16} O}
    &   \approx 638\,,
\cr
     N^{atm}_{ \nu_\mu {}^{16} O}
  &   \approx 2606\,.
\end{align}

\section{Significance}
\label{sec:significance}
In this section, we discuss the significance of both the high and low energy channels. In Ref.\cite{Ueno:2012md}, low-energy solar
neutrinos are used to bound the monopole flux $\phi_M$ to be
\begin{align}
  \label{eq:superK-lowE-bound}
  \phi_M \lesssim  6.3\times 10^{-8}\,,
\end{align}
at $90\%$ confidence level. 
Instead of going through a similar analysis with real data, we will compare
the relative signal significance of the high and low
energy channels for some fiducial values of $\phi_M$, to demonstrate that the 459 MeV antineutrinos ultimately offer better
discovery potential. 
We construct the $\chi$-square via
\begin{align}
  \chi^2
  & =
    \frac{(N_{exp} - N_{th})^2 } {\sigma^2}
    =    \frac{N^2_{\nu m}(\phi_M)}{N_{atm}}\,,
\end{align}
where $N_{\nu m}$ is the number of neutrinos from monopole catalysed processes.  We assume $N_{exp} = N_{atm}$, $N_{th}(\phi_M) = N_{\nu m}(\phi_M) + 
N_{atm}$, and that the statistical error dominates over systematic error. We
only use the total number of events for the comparison and do
not make use of the binning of the data in real experiments, as that
shape information enhances the two channels equally. 
By letting $N_{\nu m}$ be the number of low or high energy neutrinos,  we can find the signal significance of the two channels.

\begin{figure}[t]
  \centering
  \includegraphics[width=.45\textwidth]{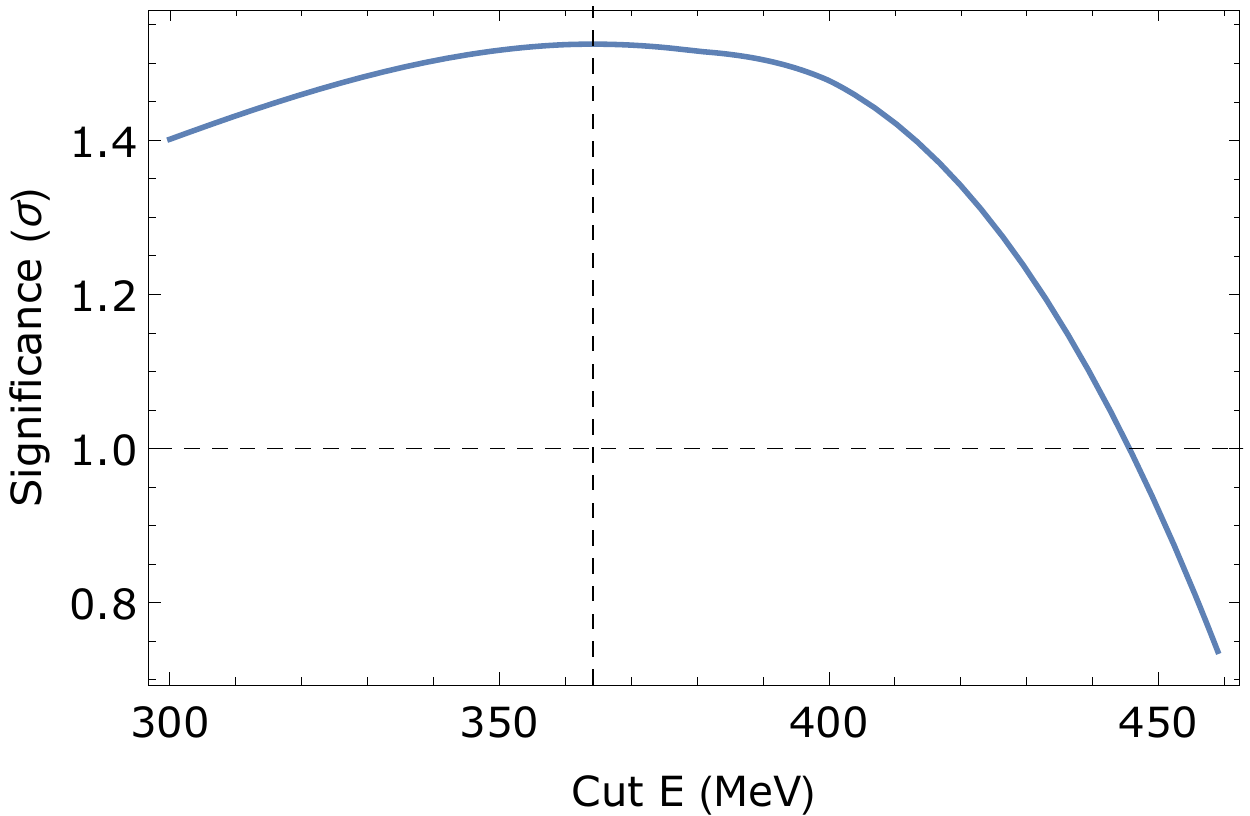}
  ~~~
  \includegraphics[width=.45\textwidth]{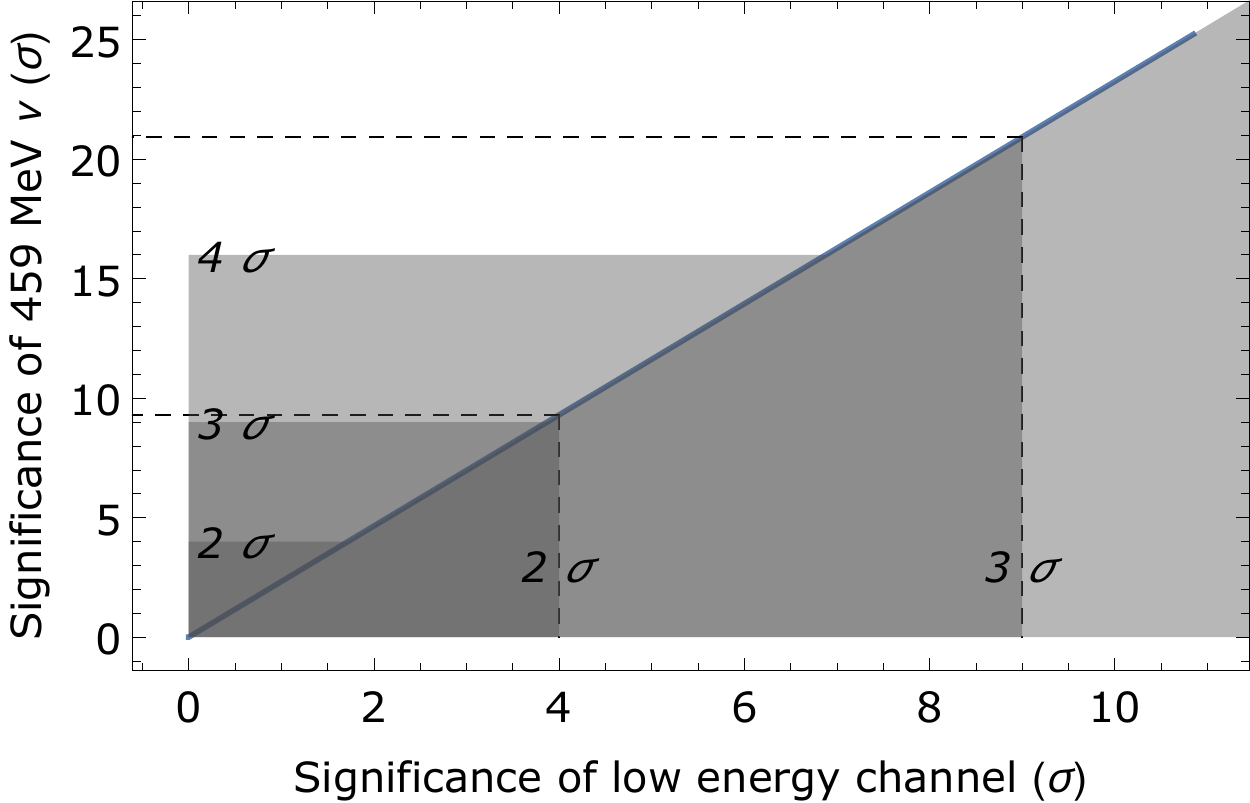}
  \caption{The statistical significance of 459 MeV antineutrino channel,
    with that of the low energy neutrino channel fixed at
    one standard deviation (left), and the comparison of the
    statistical significance of the two channels, given a fiducial monopole
    flux (right). }
  \label{fig:optimize-cut}
\end{figure}

As is shown in previous sections, the cross sections of 459 MeV
antineutrinos scattering off protons and oxygen nuclei peak at a lepton recoil energy different to that of the atmospheric neutrino
background. To maximise the significance of the 459 MeV neutrino
signal, we then float the energy cut position to maximise the $\chi^2$. In
Fig.~\ref{fig:optimize-cut} we show the significance of the 459 MeV antineutrino signal, with a
monopole flux that give a one $\sigma$ excess in the low energy neutrino channel.
We observe the optimal cut position is at $E = 364\; \mathrm{MeV}$.

\begin{table}[t]
  \centering
  \includegraphics[height=.15\textwidth]{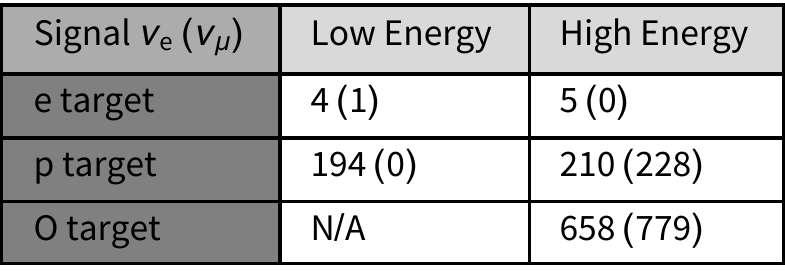}~~~~~~
  \includegraphics[height=.15\textwidth]{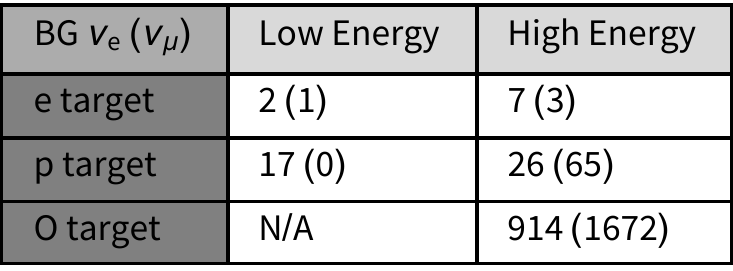}
  \caption{The signal (left) and background (right), with
    electron, proton, and oxygen target, for both channels. Please
    note that since Super-Kamiokande does not distinguish $\nu$ from $\bar
    \nu$, the background comprises both neutrinos and antineutrinos.}
  \label{tab:summary-table}
\end{table}

With this optimal energy cut, the number of events for both channels
and the background are summarised in Table~\ref{tab:summary-table}. We compare the statistical
significance of the two channels and show the result in
Fig.~\ref{fig:optimize-cut}. It is observed that for example a $2\;\sigma$ deviation
in the low energy channel will be amplified to $3\; \sigma$ in the 459 MeV
channel, and a $3\; \sigma$ effect will be amplified to more than
$4.6\;\sigma$. We also note that this is without combining the two channels
and making use of their correlation, which is likely to
enhance the result further.

\section{Discussion}
\label{sec:discussion}

As one of the most plausible aspects of physics beyond the Standard Model, magnetic monopoles have woven a persistent thread throughout particle physics for several decades. 
Whilst any hint of their discovery would be a sensation in that arena alone, it would also pose a very serious problem for inflationary theory, creating a very attractive experimental target.
Thankfully their exotic properties also allow for a range of relatively unambiguous
experimental signatures, aiding any discovery efforts. 

To this end we have examined a previously untapped discovery channel,
based on the monoenergetic 459 MeV antineutrinos produced via
monopole-induced proton decay occurring inside the Sun. 
We do note that this process relies upon the survival of the GUT monopoles carrying electroweak magnetic charge to the present day, but under the plausible assumption of a straightforward screening mechanism for these electroweak effects, this is unproblematic.
This channel was neglected in previous analyses due to the
associated electroweak suppression factor, in favour of the
unsuppressed 20 - 50 MeV neutrinos produced indirectly via
monopole-induced proton decay to neutral mesons. 

Due to the reduced experimental background and increased interaction cross section enjoyed by these high energy neutrinos, we have however demonstrated that they in fact can offer superior discovery potential.
In particular, using 5326 live days of Super-Kamiokande exposure we found that $2\;\sigma$ ($3\;\sigma$) deviations in the 20-50 MeV channel correspond to $3\;\sigma$ ($4.6\;\sigma$) deviations in the 459 MeV case. 

These effects could likely be further enhanced by leveraging the correlation between the two channels.
Liquid scintillation neutrino detectors, such as the Deep Underground Neutrino Experiment (DUNE), may also offer some distinct advantages in detecting signals of this nature and thus even greater discovery potential \cite{Kumar:2015nja, Rott:2015nma, Rott:2016mzs}.

\textit{Acknowledgments} We would like to thank Lorenzo Calibbi, Shaomin Chen, Pilar
Coloma, Tomasz Golan, Vishvas Pandey for useful communications. 
NH is supported by a CAS President's International Fellowship. 
TL acknowledges the Projects 11475238, 11647601, and 11747601 supported by the National Natural Science Foundation of China, and by the Key Research Program of Frontier Science, CAS.
CS is supported in part by the International Postdoctoral
Fellowship funded by China Postdoctoral Science Foundation, and is
grateful for the hospitality and partial support of the Department of Physics and
Astronomy at Dartmouth College where this work was undertaken.

%\nocite{*}
\bibliography{bib}

\end{document}